\documentstyle[preprint,aps,prl,floats]{revtex}

\global\firstfigfalse

\input psfig

\begin{document}

\preprint{\tighten\vbox{\hbox{\hfil CLNS 97/1479}
                        \hbox{\hfil CLEO 97-8}
}}

\title{Observation of the Decay $D_s^+\to\omega\pi^+$}

\author{CLEO Collaboration}
\date{\today}

\maketitle
\tighten

\begin{abstract} 
Using $e^+e^-$ annihilation data collected by the CLEO~II
detector at CESR, we have observed the decay 
$D_s^+ \to \omega\pi^+$.
This final state may be produced through the annihilation decay
of the $D_s^+$, or through final state
interactions. We find a branching ratio of
$\Gamma(D_s^+\to\omega\pi^+)/\Gamma(D_s^+\to\eta\pi^+)=0.16\pm0.04\pm0.03$,
where the first error is statistical and the second is systematic.

\end{abstract}

\newpage

{
\renewcommand{\thefootnote}{\fnsymbol{footnote}}

\begin{center}
R.~Balest,$^{1}$ B.~H.~Behrens,$^{1}$ K.~Cho,$^{1}$
W.~T.~Ford,$^{1}$ H.~Park,$^{1}$ P.~Rankin,$^{1}$ J.~Roy,$^{1}$
J.~G.~Smith,$^{1}$
J.~P.~Alexander,$^{2}$ C.~Bebek,$^{2}$
% ...
\end{center}
 
\small
\begin{center}
$^{1}${University of Colorado, Boulder, Colorado 80309-0390}\\
$^{2}${Cornell University, Ithaca, New York 14853}\\
% ...
\end{center}

\setcounter{footnote}{0}
}
\newpage

It has been suggested that the $\omega\pi^+$ decay mode could be
a clean signature for the annihilation decay of the $D_s^+$\cite{CHARGCONJ}.  
While the simple annihilation
diagram can produce $\rho^0\pi^+$, it
cannot produce $\omega\pi^+$, because
this final state has isospin and $G$-parity $I^G=1^+$; to do so would 
require a second-class axial current\cite{SIGNERR}. 
If at least two gluons connect the initial state quarks
to the final state quarks, the decay $D_s^+\to\omega\pi^+$ 
through the annihilation diagram is allowed.
The possibility that this final state might arise through
final state interactions (FSI) has also been extensively discussed 
\cite{KAMAL,LIPKIN,BUCCELLA}.
Fermilab E691 set a 90\% C.L.\ upper limit of
$\Gamma(D_s^+\to\omega\pi^+)/\Gamma(D_s^+\to\phi\pi^+)<0.5$\cite{E691}, or
$B(D_s^+\to\omega\pi^+)<1.8\%$\cite{PDG}; this is the most sensitive
published limit. 
To date, the only clear evidence for the annihilation decay of 
a charmed meson is $D_s^+\to\mu^+\nu$\cite{CLEOMUNU}.
This letter describes the first observation of the decay $D_s^+\to\omega\pi^+$,
and the measurement of the branching ratio 
$\Gamma(D_s^+\to\omega\pi^+)/\Gamma(D_s^+\to\eta\pi^+)$.

A recent paper by Buccella {\em et al.} 
predicts nonresonant FSI should produce
$B(D_s^+\to\omega\pi^+)=2.9\times 10^{-3}$ \cite{BUCCELLA};
however, their prediction for the related decay mode,
$D_s^+\to\eta'\rho^+$, does not agree well
with measurements\cite{PDG,CLEOETAPI}.
There could be a small
contribution to the $\omega\pi^+$ decay rate 
from spectator decay, due to the tiny $s\overline{s}$
component of the $\omega$.  The
$s\overline{s}$ content of the $\omega$ is estimated to be
$\approx0.4\%$, assuming a vector octet-singlet mixing angle of 
$39^\circ$ \cite{PDG}. The branching fraction for spectator decay to
$\omega\pi^+$ can naively be estimated to be about 
$0.004\times B(D_s^+\to\phi\pi^+)\approx 1.5\times10^{-4}$.  
This is below our current sensitivity.
There may also be  mixing of 
the $\omega$ with the $\phi$ through their common decay modes.

The data used in this analysis were collected with the CLEO~II
detector\cite{CLEONIM} at the Cornell Electron Storage Ring (CESR).  The
detector consists of a charged particle tracking system 
surrounded by an electromagnetic calorimeter.
The inner detector resides in a 
solenoidal magnet, the  coil of which
is surrounded by iron flux return instrumented with muon counters.
Charged particle identification is provided by specific ionization
($dE/dx$) measurements  in the main drift chamber.  
The data were taken at center-of-mass energies equal to the mass
of the $\Upsilon(4S)$ (10.58~GeV)
and in the continuum approximately 50~MeV below the $\Upsilon(4S)$.
The total integrated luminosity was  4.7~fb$^{-1}$.

Events used in this analysis
were required to have a minimum of three charged tracks, and
energy in the calorimeter greater than 15\% of the center-of-mass
energy. Charged tracks were
required to have $dE/dx$ measurements 
within 2.5 standard deviations
of that expected for pions.
Only energy clusters in the calorimeter with
$|\cos\theta|\leq0.71$ (where $\theta$ is the polar angle with
respect to the beam axis) 
that were not matched to charged tracks 
were used as photons.  
Photons with energy greater than 30~MeV were combined in pairs
to reconstruct $\pi^0$'s.  The invariant
mass of the two photons was required to be within 2.5~$\sigma$
of the $\pi^0$ mass, where $\sigma$ is the rms mass resolution,
about 5~MeV/$c^2$.
The $\pi^0$ candidates were kinematically fit
to the $\pi^0$ mass to improve momentum resolution; they were 
required to have a minimum momentum of 350~MeV/$c$.

To detect the decay $D_s^+\to\omega\pi^+$, we reconstructed the
$\omega$ in its dominant decay mode: $\pi^+\pi^-\pi^0$\cite{PDG}.  
We normalized to $D_s^+\to\eta\pi^+$, $\eta\to\pi^+\pi^-\pi^0$,
because it
has the same final state, so the
relative reconstruction efficiencies should be near unity
and many systematic errors cancel in the ratio.
We used  the CLEO  
Monte Carlo simulation\cite{GEANT} to determinine the ratio
of efficiencies:
${\epsilon(\omega\pi^+)/ \epsilon(\eta\pi^+)}=0.91\pm0.03$
(statistical error).  The difference from 1.00 is primarily due to 
two kinematic cuts applied to the $\omega\pi^+$ sample that were not
applied to the $\eta\pi^+$ sample (described below).

All requirements were chosen to maximize $\epsilon/\sqrt{N}$,
where the detection efficiency, $\epsilon$, 
was determined from Monte Carlo, and
the background level, $N$, from the data.  The latter was done
using $\omega\pi^+$ combinations near the $D_s^+$ mass, but
excluding a window around the $D_s^+$ mass.

We began the $\omega$ and $\eta$
reconstruction by taking pairs of oppositely charged pions, 
together with a $\pi^0$,
and calculating the invariant mass.  
Three-pion combinations whose invariant mass
was between 538 and 558~MeV/$c^2$ ($\pm2\sigma$ around the $\eta$ mass)
were used as $\eta$ candidates.
Combinations with invariant mass
between 762 and 802~MeV/$c^2$ were used as $\omega$ candidates;
this is about a $\pm0.9$~FWHM cut around the $\omega$ mass
The $\omega$ line shape is the convolution of its natural width
($\Gamma=8.4$~MeV/$c^2$ \cite{PDG}) and the detector resolution 
($\sigma\approx8$~MeV/$c^2$).

The $\eta$ and $\omega$ candidates were combined with a charged 
pion to form $D_s^+$ candidates. 
The three charged tracks, two from the $\eta$ or
$\omega$, along with this ``bachelor'' pion, were required to be 
consistent with coming from a common vertex.  
The tracks were refit to pass through this vertex, which improves the
$D_s^+$ mass resolution by about 4\%.

To take advantage of the hard fragmentation of continuum charm, 
we required
all $D_s^+$ candidates to have $x\geq0.6$, 
where $x$ is the scaled momentum:  $x\equiv p/p_{max}$ and
$p_{max}=(E_{beam}^2 - {M_{D_s^+}}^2)^{1/2}$.  This suppresses
combinatoric background. A cut on the decay angle
of the $D_s^+$ was also applied.  The decay angle ($\theta_\pi$)
is defined as the
angle between the bachelor pion in the $D_s^+$ rest frame, and the
$D_s^+$ momentum in the lab frame.  Since the $D_s^+$ has $J=0$, the decay
angle must have a flat distribution for the signal, while the background 
peaks toward $\cos\theta_\pi=-1$.  A cut of $\cos\theta_\pi\geq-0.85$ 
was used; this retains 92\% of the signal and 60\% of the background.

Two kinematic cuts were applied to the $\omega\pi^+$
combinations.  First, because the $\omega$ is a vector particle,
it must be produced in the helicity-zero state in the decay 
$D_s^+\to\omega\pi^+$.  We define the helicity angle, $\alpha$, to
be the angle between the normal to the $\omega$ decay plane and
the $D_s^+$ direction, both measured in the $\omega$ rest frame.  This
angle must have a distribution proportional to $\cos^2\alpha$.  We required
$|\cos\alpha|\geq0.45$.  This cut keeps more than 90\% of the
signal and about 55\% of the background. 

Second, the amplitude for the $\omega$ decay is maximal at the center
of the Dalitz plot.  We calculated a parameter which is proportional
to this decay amplitude; it is simply the cross-product of two
of the pions' momenta, measured in the $\omega$ rest frame.  The 
parameter ($R$) was normalized so that it equals one at the center of
the Dalitz plot, and goes to zero at the edge.  We required
$R^2\geq0.2$; this retains 97\% of the signal and about 80\% of the
background.

\begin{figure}[t]
\centerline{\psfig{figure=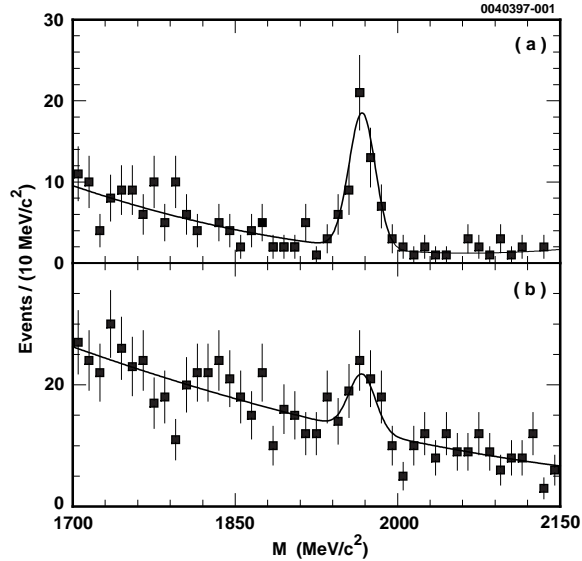,height=3.in}}
\smallskip
\caption[]{
Histogram of (a) $\eta\pi^+$, and (b) $\omega\pi^+$ invariant mass for 
tagged events.
The points with error bars are the data; the solid
lines are the result of the
constrained fit to the data, as described in the text.} 
\label{fig1}
\end{figure}

Finally, we sorted the $D_s^+$ candidates into two categories:
``tagged'' and ``untagged''.  The tagged events are those that are
consistent with coming from the decay $D^{*+}_s\to D_s^+\gamma$. To 
tag events, we 
combined the $D_s^+$ candidates with photons and calculated
the invariant mass of each $D_s^+\gamma$ combination.
To suppress mistags from energy clusters produced by hadronic interactions,
we required   
the tagging photon's energy be at least 250~MeV
and its lateral shape to be consistent with an electromagnetic shower.
We calculated 
the mass difference $\Delta M_\gamma\equiv M(D_s^+\gamma) - M(D_s^+)$.  
The $D_s^+$ is ``tagged'' if 134~MeV/$c^2\leq\Delta M_\gamma<154$~MeV/$c^2$. 
Events in which no photon meets 
this criterion are ``untagged''.
The invariant mass distribution of the tagged $\eta\pi^+$
combinations is shown in Fig.~1a.  The histogram has been fit with
a Gaussian for the $D_s^+\to\eta\pi^+$ events 
and a second-order polynomial for the combinatoric background.   
The mean and width of the Gaussian were fixed to the values
predicted by Monte Carlo.
The fit finds $48.4^{+8.4}_{-7.7}$ signal
events (statistical error only). 
The overlayed functions shown in the figure are 
the result of a more constrained fit described below.
About 3\% of the events contained more than one $\eta\pi^+$
combination which satisfied our criteria.  The same is true in the
$\omega\pi^+$ mode.  Since this occurred at the same rate in the
data and Monte Carlo, and in both the signal and normalizing modes,
we accepted these double-counts; they have negligible effect on our
results.

A histogram of the invariant mass of the
tagged $\omega\pi^+$ combinations is shown in
Fig.~1b.  It was fit with the same functions as the $\eta\pi^+$ data,
using the same Gaussian parameters, as predicted by the Monte Carlo.
This fit finds $35.7^{+10.8}_{-10.2}\ D_s^+\to\omega\pi^+$ 
events (statistical error only). We consider this to be a significant
signal and describe further tests of the data below.  

A number of checks have been performed to help validate this signal.
Three-pion combinations were selected in sidebands to the $\omega$
signal region:  670~MeV/$c^2 \leq M(\pi^+\pi^-\pi^0)<710$~MeV/$c^2$ and
855~MeV/$c^2\leq M(\pi^+\pi^-\pi^0)<895$~MeV/$c^2$.  When these are
combined with a fourth pion, and the $\omega\pi^+$ selection 
criteria applied, no $D_s^+$ signal is seen in either sideband.
To reproduce the observed $\omega\pi^+$ signal would
require a 6 standard deviation fluctuation.

One can also fit the $\Delta M_\gamma$ distributions for a signal.  
Requiring that the four-pion ($\eta\pi^+$ or $\omega\pi^+$) mass 
be between 1943 and 1991~MeV/$c^2$\ 
and removing the cut on $\Delta M_\gamma$, 
we found $50^{+10}_{-9}$ $\eta\pi^+$ events and
$42^{+14}_{-13}$ $\omega\pi^+$ events, 
in good agreement with the yields found in
the previous fits to the $\eta\pi^+$ and $\omega\pi^+$ mass
distributions (Fig.~2). In these $\Delta M_\gamma$
histograms, double-counting occurred at a rate of about 10\%;
this is neglible compared to the statistical errors.

\begin{figure}[t]
\centerline{\psfig{figure=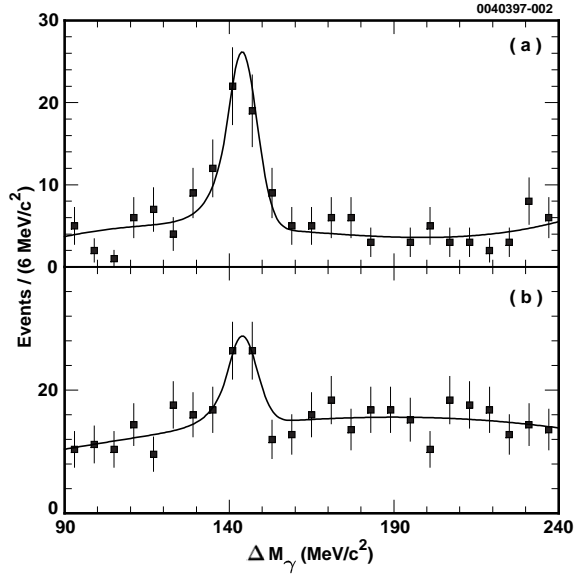,width=3. in}}
\smallskip
\caption[]{
Histogram of $\Delta M_\gamma$ for (a) $\eta\pi^+$ events, and (b)
$\omega\pi^+$ events.
The points with error bars are the data. The solid lines are fits,
using a modified Gaussian for the signal, whose shape was fixed using
Monte Carlo events, and a third-order polynomial for the
background.} 
\label{fig2}
\end{figure}

To confirm that these events are in fact $D_s^+\to\omega\pi^+$, rather
than some other four-pion decay of the $D_s^+$, we 
loosened the $\omega$ mass cut and took all $\pi^+\pi^-\pi^0$
combinations with masses between 650 and 900~MeV/$c^2$.  These were then 
combined with a fourth pion; the four-pion combinations that passed
the tagging criteria (and all other cuts) were kept.  
Again requiring that the four-pion mass 
be between 1943 and 1991~MeV/$c^2$, we made a histogram of the three-pion
invariant mass.    A fit to this histogram 
yields $44\pm12$ events.  However, there are also real $\omega$'s
in the $\omega\pi^+$ random combinations under the $D_s^+$ peak.
To account for this, we performed a sideband subtraction, using
upper and lower sidebands in four-pion mass.
After the subtraction, a fit to the
three-pion invariant mass  found $32\pm12$
$\omega$'s, consistent with our previous results.

We have calculated the invariant mass of the ``other'' three
pion combination in each $\omega\pi^+$ candidate event.  
We define ${M_3}'$ to be the invariant mass of the 
bachelor $\pi^+$ with the $\pi^-$
and $\pi^0$ from the $\omega$.  For the $\omega\pi^+$ events, all of 
the events in the $D_s^+$ signal region 
have ${M_3}'>1100$~MeV/$c^2$.  
Thus these events are not simply $D_s^+\to\eta\pi^+$
or $D_s^+\to\phi\pi^+$ (with $\phi\to\pi^+\pi^-\pi^0$) events feeding into
$\omega\pi^+$ by combining the pions in the ``wrong'' order. The
${M_3}'$ distribution agrees with the signal Monte Carlo prediction.

Similarly, we reconstructed events in the signal region as
$K^-\pi^+\pi^+\pi^0$, as might come from $D^{*+}$ decay, by
assigning the kaon mass to the negatively-charged track. We found
that the invariant mass for this alternate particle assignment
in every case is more than 2040~MeV/$c^2$, so these cannot be 
misreconstructed $D^{*+}$ events.
Again, the measured distribution agrees with the Monte Carlo prediction.

The untagged sample of $\omega\pi^+$ events 
contains a small excess at the $D_s^+$ mass (Fig.~3).  A fit yields $133\pm57$
signal events.  Fitting the untagged $\eta\pi^+$ distribution
finds $312\pm31$ signal events.
We included these untagged events in the branching ratio measurement.

\begin{figure}[t]
\centerline{\psfig{figure=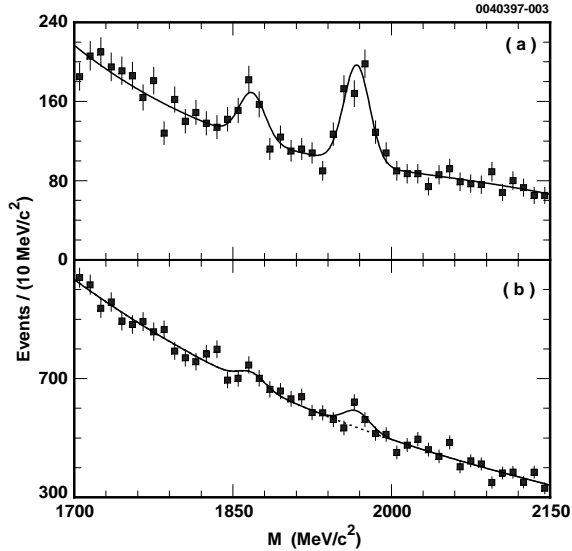,width=3. in}}
\smallskip
\caption[]{
Histogram of (a) $\eta\pi^+$, and (b) $\omega\pi^+$ invariant mass for 
untagged events.
The points with error bars are the data; the solid
lines are the result of the
constrained fit to the data, as described in the text. The fits include
a Gaussian of fixed mean and width for $D^+$ events near 1869~MeV/$c^2$.
In the lower plot, the dashed line shows the background function
underneath the $D_s^+$ signal; the $y$-axis scale has been zero-suppressed.} 
\label{fig3}
\end{figure}

The ratio of reconstruction efficiencies, 
$\epsilon(\omega\pi^+)/\epsilon(\eta\pi^+)$,
is the same for tagged and untagged events,
so the raw ratio of signal events should also be the same in both
samples.  For the tagged events, we find a ratio of $0.74\pm0.25$
\ $\omega\pi^+$ event per $\eta\pi^+$ event.  For the untagged
events, the ratio is $0.43\pm0.19$.  The two ratios are
statistically consistent.

We also performed a simultaneous fit to the four
distributions ($\eta\pi^+$ and $\omega\pi^+$, tagged and untagged),
and constrained  the ratio of $\omega\pi^+$
to $\eta\pi^+$ events to be the same for both 
samples.  This yielded a ratio of $0.56^{+0.15}_{-0.14}$; the
$\chi^2$ of the fit to the four histograms
was 146.8 with 161 degrees of freedom.  We used
the result of this constrained fit in the branching ratio calculation;
the fit functions shown in figures 1 and 3 are also the result of this
fit.  Refitting the histograms with the number of $D_s^+\to\omega\pi^+$
events fixed to be zero yielded a $\chi^2$ of 166.9, an increase of 20.1.

Using the ratio of efficiencies determined from Monte Carlo and the
$\eta$ and $\omega$ branching fractions to
$\pi^+\pi^-\pi^0$\cite{PDG}, we determined the branching ratio:
\begin{equation}
{\Gamma(D_s^+\to\omega\pi^+)\over \Gamma(D_s^+\to\eta\pi^+)}=
                     0.16\pm0.04\pm0.03. \label{ratio}
\end{equation}
The first error is statistical; the systematic error is dominated
by variations in the branching ratio caused by varying the cuts
used in the analysis. These variations help gauge the accuracy of our
event simulation. The systematic error also includes 
contributions from the uncertainty in the efficiencies, the branching
fractions of the $\eta$ and $\omega$, and from variations in the
result using different fitting functions.

In order to calculate an absolute branching fraction for 
$D_s^+\to\omega\pi^+$, we used the new CLEO measurement 
${\Gamma(D_s^+\to\eta\pi^+) / \Gamma(D_s^+\to\phi\pi^+)}
=0.47\pm0.07\cite{CLEOETAPI},$
and the PDG value of $B(D_s^+\to\phi\pi^+)=0.036\pm0.009$\cite{PDG}.  This
yields a branching fraction:
\begin{equation}
B(D_s^+\to\omega\pi^+)=(2.7\pm1.2)\times10^{-3}, \label{fraction}
\end{equation}
where all the errors have been added in quadrature. 
Thus we have observed
the decay $D_s^+\to\omega\pi^+$, which
may be the result of annihilation decay, 
final state interactions, or both.

We gratefully acknowledge the effort of the CESR staff in
providing us with excellent luminosity and running conditions.
This work was supported by the National Science Foundation, the
U.S.\ Department of Energy, the Heisenberg Foundation, the 
Alexander von Humboldt Stiftung, Research Corporation, the
Natural Sciences and Engineering Research Council of Canada,
and the A.P.\ Sloan Foundation.

\end{document}